# Nanoscale Fabrication by Intrinsic Suppression of Proximity-Electron Exposures and General Considerations for Easy & Effective Top-Down Fabrication

H Bartolf[*], K Inderbitzin, L B Gómez, A Engel, A Schilling

Physics-Institute, University of Zürich, Winterthurerstrasse 190, 8057 Zürich, Switzerland

*E-mail: holger.bartolf@physik.uzh.ch, +41 44 635 5722

**Abstract** We present results of a planar process development based on the combination of electron-beam lithography and dry etching for fabricating high-quality superconducting photosensitive structures in the $sub-100$nm regime. The devices were fabricated by the application of an intrinsic proximity effect suppression procedure which makes the need for an elaborated correction algorithm redundant for planar design layouts which are orders of magnitude smaller than the backscattering length. In addition, we discuss the necessary considerations for extending the fabrication spatial scale of optical contact-lithography with a mercury arc-discharge photon source down to the order of the exposure photon's wavelength ($sub-\mu m$), thereby minimizing the writing time on the electron-beam lithograph. Finally we developed a unique and novel technique for controlling the undercut during a planar lift-off fabrication procedure without cleaving the wafer.

# 1. Introduction

Top-down fabrication of mechanically and/or electronically contactable structures with a spatial resolution on the $sub-100nm$ scale is very often achieved by combining the technological approaches of photolithography [1] and electron-beam lithography (EBL) [2,3]. The seminal character of this approach is based on mixing the fast, but inflexible light-patterning technique with the relatively slow, but very flexible patterning technique that uses a deflectable, focused beam of electrons [4,5]. This so-called mix and match technique serves as a powerful tool for easy, effective and flexible nanoscale device fabrication [6].

For most research and development (R&D) EBL projects, a relatively high acceleration voltage of about $30kV$ is used to reduce the forward scattering angle (see chapter 3.3.3 in [4]) and hence to increase the achievable resolution (R&D relies on a relatively low-cost accessory, so that we restrict the discussion here to such systems). From a fabrication engineer's point of view, the drawback of such a high tension is the increase of elastic large-angle backscattering events of electrons (see chapter 4.1 in [4]) that come close to the nuclei of the atomic species of the substrate. These backscattering events are characteristic for different substrate materials and are associated, per definition, with energies larger than $50eV$ (see figure 1.5 in the introductorily chapter of [4] for a vivid comparison of the different electron energies during the electron-matter interaction). Therefore, the backscattered electrons induce an exposure on a radially symmetric scale of a few microns around the exposed pixel by generating low-energy (on the same order as the binding forces within a molecule) secondary electrons [7] (see also chapter 4.2 in [4]) which induce the lithographically necessary chemical reaction in the electron-sensitive organic resist that is spin-coated on top of the sample. This unintended exposure of secondaries generated by backscattered electrons is called *proximity effect in electron-beam lithography* (title of [8]). Therefore, EBL fabrication length scales on the order or smaller than the backscattering length require complex and time-consuming algorithms [9,10] that correct for this proximity exposures. The focus of this paper discusses a unique, novel and simple EBL design procedure that makes corrections for proximity exposures almost needless. Our approach can be applied to device layouts which have planar spatial dimensions that are significantly smaller than the range of the backscattered electrons. This requirement is in particular fulfilled for high acceleration voltages and for fabrication length scales in the $\text{sub}-100\text{nm}$ regime, which is two orders of magnitude smaller than the range of the backscattered electrons in silicon or sapphire at $30\text{kV}$ (see figure 2.4 in [11] for the Bethe-range plotted over the accelerating voltage and chapter 4.1.4 in [4] for the angular distribution of backscattered electrons for different materials).

On the other hand, the mix and match planar top-down fabrication approach requires an accurate alignment of the micro- with respect to the nanolithographically generated pattern. Therefore, the electromechanical motion of the piezoelectronically controlled laser-interferometer stage and the electron-beam deflection unit needs to be aligned precisely by scanning a set of global and local markers. Such an automatic marker recognition procedure on pre-patterned wafers allows for mix and match processes with extremely high overlay accuracy in the $\text{sub}-20\text{nm}$ regime [12]. For most R&D projects, contact photolithography, physical vapour deposition and a standard lift-off technique are used to supply these global and local markers. When using this patterning technique it is advantageous to fabricate even micro-scaled parts of the latter device with contact photolithography in order to reduce the writing time of the EBL. Hence it is desirable to achieve the ultimate resolution during contact photolithography, which can be pushed down to the order of the wavelength of the exposure-photons spontaneously emitted during the discharging arc-glow [13] (smaller than $500\text{nm}$ in case of a mercury plasma discharge). Therefore, the second part of this work discusses the limitations and necessary considerations when shrinking the critical dimensions of the light-patterned structures into the $sub-\mu m$ scale by employing a lift-off deposition process using an image reversal resist. Furthermore, a generally applicable innovative approach is presented that allows for a comfortable control over the undercut necessary for the lift-off procedure during an additive lithographic device fabrication process step.

This article is structured as follows: First, we discuss a micromechanical top-down fabrication approach that intrinsically suppresses proximity exposures without the need for a sophisticated correction algorithm. It is based on the correct choice of positive or negative electron-sensitive resist for the lithographic pattern transfer. In chapter 3, we present an approach to easily obtain high-quality photolithographically defined pattern with critical fabrication dimensions very close to the wavelength of the photons emitted by the light source. Then we describe a novel technique for controlling the undercut during the chemical development without cleaving the wafer. Finally, we shortly discuss a complete elaborated process design-layout that demonstrates the highly desirable flexibility of a mix and match approach that allows for easy & effective fabrication of many devices per fabrication run. Throughout this work, a Raith150 EBL system was used.

## 2. Intrinsic Proximity Effect Suppression during Electron-Beam Lithography

The figures and examples of the here presented nanoscale approach (see chapters 2.2.1 and 2.2.2) are mainly based on the layout of superconducting nanowire highspeed single-photon detectors [14-17] for the visible and near infrared region of the electromagnetic spectrum (see figures 2, 5). Such detectors find, among other things, applications in the emerging field of quantum cryptography [18-20].

However, the applicability of our approach developed within this chapter can be extended to any design, as long as the desired pattern is significantly smaller than the backscattering length as will be discussed in addition in chapter 2.2.3 for the design of a bridge and a SQUID.

*2.1 Proximity Effect Model(s)*

The general phenomenological approach to mathematically model the proximity exposure discussed in the introduction due to beam-broadening and backscattering can be visualized, for example, in Monte Carlo simulations (for an early work consider figure 1 in [21], and [22] for a modern discussion) and is modelled by the proximity function

$$f_{\text{Prox}}(r) = \frac{1}{\pi(1+\eta)}\left[\frac{1}{\alpha^2}\exp\left(-\frac{r^2}{\alpha^2}\right) + \frac{\eta}{\beta^2}\exp\left(-\frac{r^2}{\beta^2}\right)\right]; \quad \alpha \ll \beta; \quad \alpha, \beta > 0 \qquad (1)$$

which describes the deposited energy induced by forward and backward scattered electrons by a superposition of two gaussian functions in planar polar coordinates (defined by the radial and angular coordinates $r$ and $\varphi$) around the exposed pixel located at $r = 0$. The sum of these two contributions for the marginal case $r \gg \beta$, that "clears" the exposed area during the chemical development procedure, is called the clearing (area) dose value and is measured in $\mu C/cm^2$. The exposure originated by forward scattered electrons takes place in the range of the forward scattering length $\alpha$. With the commonly known contamination dot technique, we determined $\alpha = 5\,\text{nm}$ (**f**ull **w**idth at **h**alf **m**aximum FWHM=$8.33\,\text{nm}@30\,\text{kV}$, see inset of figure *Contamination Dot* in part I of [23]). The exposure due to backscattered electrons is determined by the backscattering length $\beta$ and the backscatter coefficient $\eta$. The backscatter coefficient determines how much energy is deposited in the resist per incident electron energy. The phenomenological parameters $\beta$ and $\eta$ depend on the density and the atomic number of the elements of the substrate material.

The prefactor $(\pi(1+\eta))^{-1}$ ensures the normalization of the proximity function to unity. For sapphire substrates, which are used for fabricating our nanowire highspeed detectors (see chapter 2.3), the proximity parameters are $\beta=2.436\,\mu m$ and $\eta=0.5$ (after [24]). In order to visualize the influence of the proximity parameters, equation (1) is plotted for several different $\eta$ values but constant $\beta=2.436\,\mu m$ in figure 1.

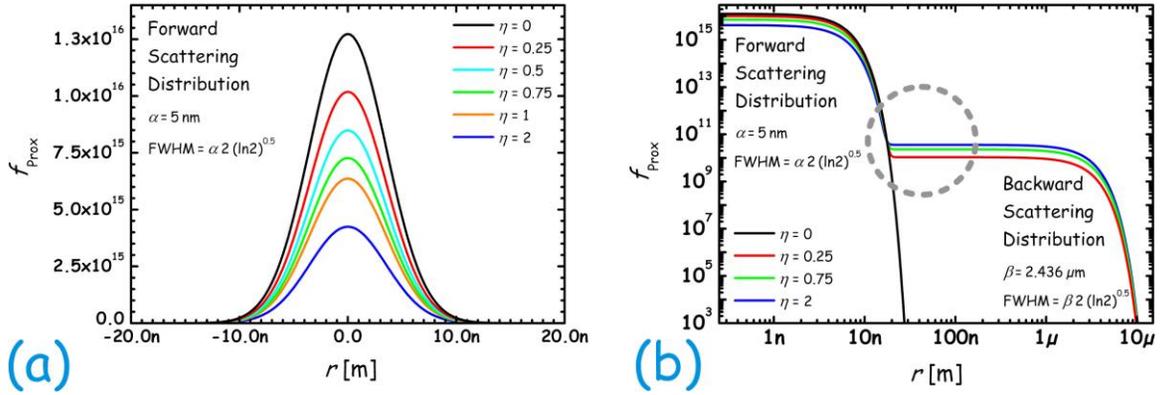

Figure 1: Picture (a) shows the forward scattering Gaussian distribution of the focused electron beam. Picture (b) displays the proximity function (equation (1)) in a log-log plot ($\alpha=5\,nm$ and $\beta=2.436\,\mu m$). The midrange of the scattered electrons is coarsely visualized by a dashed circle where the Gaussian model given by equation (1) oversimplifies the electron-matter interaction. In this midrange the model of equation (1) is extended by a third (exponential) term (equation (2)). Details are explained in the text.

Figure 1(a) shows the forward scattering distribution of the focused electron-beam. Figure 1(b) displays the contribution from both scattering effects in a log-log plot. With increasing backscattering (growing $\eta$), the weight of the exposure by forward scattered electrons is reduced in favour of the backscattering contribution. This effect is visualized in figure 1(b). The exposure induced by backscattering is normally about six orders of magnitude smaller than the contribution by forward scattering. However, it occurs on a much larger spatial scale. As a direct consequence, the deposited dose of many narrowly spaced pixels (we choose a pixel separation $\Delta x=10\,nm$) sums up the dose-contribution from backscattered electrons into a range that is comparable to the induced exposure due to forward scattering. If not corrected properly, this contribution from backscattered electrons can lead to unintended intra- and intershape proximity exposures (see figures 7a and 8a in [25] and figure *Clearing Dose Determination* in part I of [23] for vivid examples of intra- and intershape proximity exposures).

For instance, in case of an exposed square with an area $a^2$ that is orders of magnitude smaller than the square of the backscattering length $\beta^2$, the second term in equation (1) can be neglected for calculating the deposited dose, resulting in $\int_0^{2\pi} d\varphi \int_0^{\infty} f_{\text{Prox}}(r) r dr = (1+\eta)^{-1}$. Consequently such a tiny area on a sapphire substrate has to be exposed by 1.5 times the clearing dose, because it misses completely the dose contribution from the backscattered electrons. This missing dose is compensated by a 50% longer exposure time. On the other hand, a large squared area that is orders of magnitude larger than $\beta^2$, does not require a correction for proximity exposures and is therefore exposed by depositing the clearing dose value 1.0.

The model given by equation (1) is somewhat oversimplified, but it is sufficient to understand the basic principle of the proximity effect and the established approach for its correction. For proximity correction, one has to determine the values of $\alpha$, $\beta$ and $\eta$ experimentally e.g. using the so-called doughnut-structure method [26-28]. Once the parameters of the proximity function are determined, the polygons to be exposed by the electron beam are fragmented into smaller rectangles and/or triangles by a mathematical algorithm depending on $\alpha$, $\beta$ and $\eta$ (see upper row of figure 2). The intention is to assign multiples of the predetermined clearing dose (in the range from 1.0 to 1.5 in case of sapphire) to the smaller shapes in such a way that the clearing dose value is homogenously distributed over the exposed area(s). In other words the sum of forward scattered and backscattered electrons is equally distributed within the exposed area after the application of the correction algorithm. This sum is equal to the clearing dose value.

More accurate models [29,30] add a third exponential summand term

$$\frac{\nu}{2\gamma^2} \exp\left(-\frac{r}{\gamma^2}\right) \qquad (2)$$

to equation (1) to account for the midrange energy deposition. Now a factor $(\pi(1+\eta+\gamma))^{-1}$ ensures the normalization of the proximity function $f_{\text{Prox}}$ to unity. The midrange is coarsely indicated by a dashed grey circle in figure 1(b). In this extended model, two additional parameters $\nu$ and $\gamma$, representing the strength and range of midrange exposure, have to be determined again e.g. with the doughnut-structure method [26-28]. However, accounting for midrange exposures will only slightly influence the dose values of the smaller shapes that build up the structure to be exposed.

A powerful proximity effect correction software package (NanoPECS[TM]; **P**roximity **E**ffect **C**orrection **S**oftware) [9,10] is implemented within the software suite used on the Raith150. It allows for an accurate proximity exposure correction of arbitrary structures if one has experimentally determined the range and strength of the exposures induced by the different scattering mechanisms. We used this software package to demonstrate that elaborated corrections for the proximity effect become unnecessary for individual device layouts with lateral dimensions that are significantly smaller than the backscattering length $\beta$. When using NanoPECS[TM], we ignored the midrange exponential term given by equation (2) by setting $\nu=0$ which leads directly to equation (1). We will discuss our approach in the following section and present the obtained results in section 2.3.

*2.2 Simulated Proximity-Effect Correction*

For fabricating devices from a deposited thin film using a subtractive lithographic combination of EBL and dry etching in a reactive plasma discharge, one has in general two possibilities. In case of a positive resist, the electron beam has to be guided around the desired structure to generate the resist protection for the etching during the chemical development. In the other case of negative resist, the beam has to be guided over the complementary design. For a device pattern that has much smaller lateral planar dimensions than the backscattering length $\beta$ these two possible approaches show an entirely different character when corrected for the proximity effect.

*2.2.1 Correction for Proximity Exposures for the Design of a Nanowire Single-Photon Detector*

Using the simple double Gaussian of equation (1) and the proximity parameters ($\beta = 2.436\,\mu\text{m}$, $\eta = 0.5$) determined by [24], the correction for exposures induced by backscattered electrons on a sapphire substrate was calculated with NanoPECS$^{\text{TM}}$ for the design of a nanowire detector structure ([14-17]). The resulting fracturing by NanoPECS$^{\text{TM}}$ is shown in figure 2 for a designed nanowire path-width of $w = 90\,\text{nm}$. Our design allows for four-probe electronic-transport characterization of the subsequently fabricated device. The single-pixel detection element consists of a superconducting conduction path of width $w$ that covers the detection area $A_{\text{detec}} = l_1 \cdot l_2$ in form of a meander. The conduction paths are connected by islands with an area $a_{\text{I}} \cdot b_{\text{I}}$ where $b_{\text{I}} = 2a_{\text{I}}$ is set for scaling purposes. The islands were placed to avoid a possible constriction while the conduction path turns around by $180°$. The spacing $s$ of the conduction paths determines the detector's filling factor $FF$. For practical reasons, a high $FF$ and hence small $s$ is desirable for detector applications as will be discussed in section 2.3. For an infinite detection area, the filling factor is defined as $FF_\infty = w/(w+s)$. Throughout this paper, the exact definition for a finite detection area $FF = N_p w l_2 / A_{detec}$ is used, where $N_p$ is the number of conduction paths ($FF \xrightarrow{N_p \to \infty} FF_\infty$).

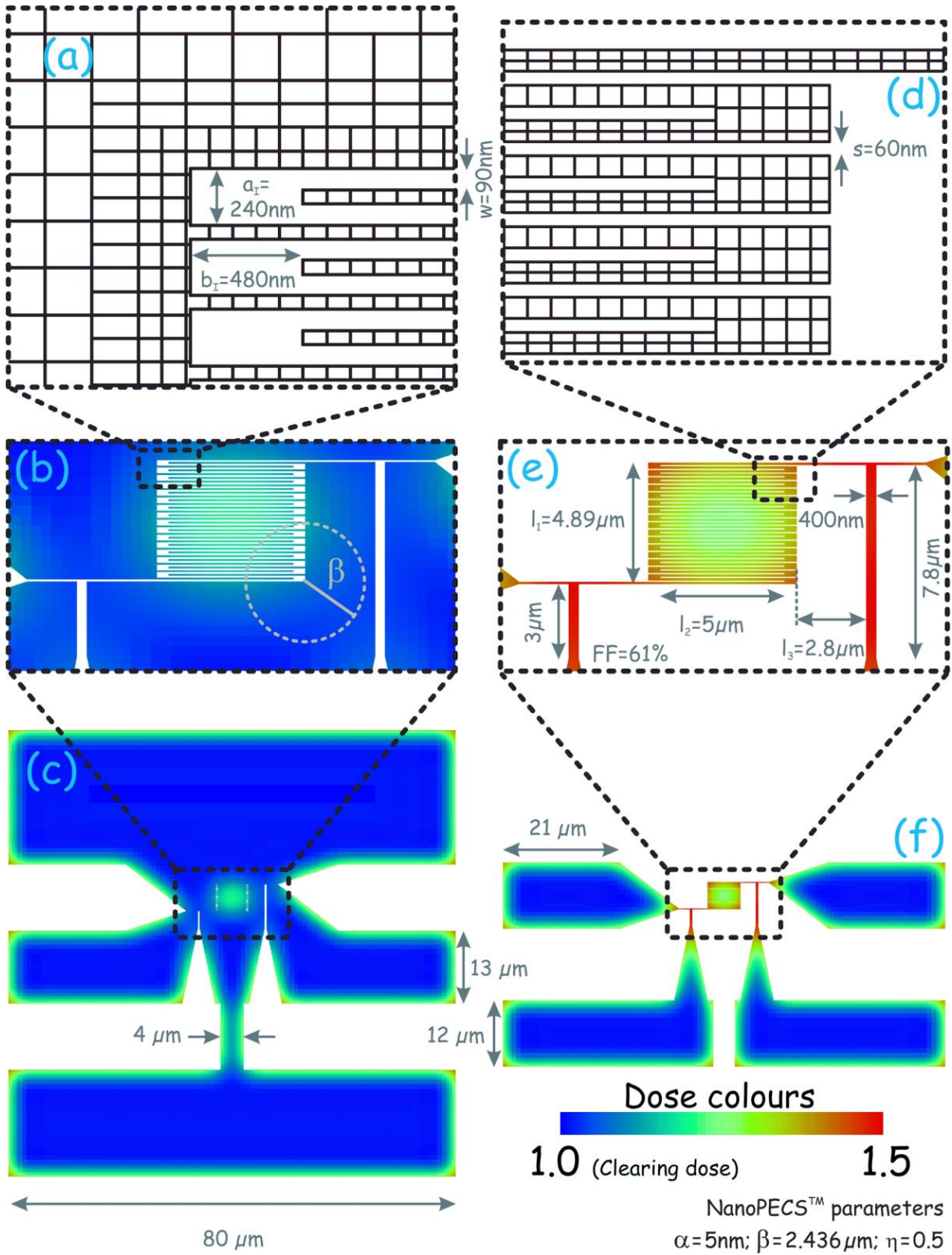

Figure 2: Correction for the proximity effect on a sapphire substrate for exposures around the desired structure ((a)-(c), left) and exposures of the desired structure ((d)-(f), right). The whole structure (single-photon detector with $w=90\text{nm}$) is fragmented into smaller primitive rectangular structures (shown in the top row (a), (d)) by NanoPECS™. The colours indicate the dose of these primitives in multiples of the clearing dose value 1.0. The exposure around the structure ((a)-(c), left) needs virtually no correction for device dimensions which are significantly smaller than the backscattering length $\beta = 2.436\,\mu\text{m}$. For this particular meander design $N_\text{p} = 33$.

Analogous to the design of figure 2 we investigated four different conduction path widths ($w=90\text{nm}, 120\text{nm}, 200\text{nm}$ and $300\text{nm}$) that were separated by the minimum design value $s=60\text{nm}$. A smaller design path spacing leads to merging of the paths after the etching (see section 2.3). In case of the two smaller conduction path widths the detection area was $\approx 5\,\mu\text{m} \times 5\,\mu\text{m}$. In the other two cases, the path covered an area of about $10\,\mu\text{m} \times 10\,\mu\text{m}$. In the direction parallel to the current path, the detection area length $l_2$ was exactly $10\,\mu\text{m}$ (or $5\,\mu\text{m}$, respectively) while in the direction perpendicular to it the number of meander turns was chosen in a way that the detection area width $l_1$ was slightly less than $10\,\mu\text{m}$ (or, $5\,\mu\text{m}$ respectively). The corresponding filling factors in these designs were $FF=61\%, 68\%, 77\%$ and $84\%$. Within our layouts (analogous to figure 2), the spacing of the voltage and current leads with respect to the nanoscale device structure was chosen to be larger than the backscattering length (in particular $l_3 = 2.8\,\mu\text{m} > \beta$, see figure 2(e)). The EBL design shown in figure 2 fits into a write field of $(100\,\mu\text{m})^2$ for the focused electron-beam. Therefore the writing times for exposures of the desired structure and exposures around the desired structure are comparable to each other. The writing time for exposures around the meander structure is slightly longer and is determined by the beam current $I_\text{B}$, the pixel separation $\Delta x$ and the beam speed $v_\text{B}$, because the locally deposited exposure dose $D_\text{exp}$ is given by

$D_{\text{exp}} = I_B / (\Delta x \cdot v_B)$ in µC/cm$^2$. The beam speed is adapted from one smaller primitive to another to allow for the deposition of different dose values measured in multiples of the clearing dose.

*2.2.2 Discussion about the Advantage of Intrinsic Proximity Exposure Suppression*

In figure 2 the coloured areas represent the deposited dose of the simulated electron-beam exposure. In the first case (negative etch mask, right part of figure 2(d)-(f)), the beam is guided directly over the meanders' area. The conduction paths, the voltage and current leads are much smaller than the backscattering length. Therefore these structures miss a significant contribution from backside exposure. Consequently, they need a higher dose (about $(1+\eta) = 1.5$ times the clearing dose value). On the other hand, if the beam is guided around the desired structure the exposed pattern is practically undisturbed by proximity effect corrections (positive etch mask, left part of figure 2(a)-(c)). The reason for this is that the exposed pattern is interrupted only on a tiny length scale that is much smaller than the backscattering length $\beta$, which is true for all of the above mentioned investigated path widths. As a direct consequence of this model investigation, an exposure *around* a desired structure without the usage of the NanoPECS$^{\text{TM}}$ fracturing algorithm is preferable due to its *intrinsic* proximity-exposure suppression. See chapter 2.2.3 for a deeper discussion of this approach for the design of a bridge and a **s**uperconducting **qu**antum **in**terference **d**evice SQUID [31,32].

The advantage of this intrinsic suppression of exposures induced by backscattered electrons is obvious: The phenomenological proximity parameters $\alpha$, $\beta$ and $\eta$ do not have to be determined experimentally. Consequently, the same structure can be easily "*cloned*" by changing the substrate material, substrate density and/or the acceleration voltage as long as it is assured that the backscattering length is still orders of magnitude larger than the critical device dimensions, which can be coarsely checked with a simple Monte Carlo simulation algorithm (e.g. CASINO [33-35]).

A further advantage of our approach is that the complexity of a nanoscale fabrication process is reduced and time- and/or computational-power consuming proximity-effect calculations are no longer needed.

Furthermore, the abandonment of the NanoPECS$^{TM}$-fracturing algorithm allows for guiding the electron-beam parallel along the conduction paths (x-direction in figure 3) without any interruption by the deflection coils of the 16-bit pattern generator of the electron-beam lithograph which is sketched schematically in figure 3 for the layout of a simple bridge. Hence spatial patterning uncertainties due to repositioning the beam from one primitive structure to the next one do not play a role any more. In addition one observes less dynamic effects that originate from the finite inductance of the lithograph's deflection coil system and that can occur during the acceleration or deceleration of the beam to the beam speed $v_B$ at the borders of the primitive structures. In order to diminish these dynamic effects, we have chosen the smallest aperture on the Raith150 which resulted in a beam-current of $27\,\text{pA}$. To deposit the clearing dose value of $55\,\mu\text{C/cm}^2$ (see chapter 2.3) with the used pixel separation $\Delta x = 10\,\text{nm}$ a beam speed of $v_B = 4.9\,\text{mm/s}$ is necessary. For the design drawn in figure 2(c), exposed analogous to the mode sketched in figure 3, the focused electron-beam of the Raith150 needs about $2\,\text{min}$.

Finally, the choice of positive resist for the etch mask has the further advantage that the $30\,\text{kV}$-electrons of the beam do not interact with the latter device structure. At this point, one might argue that the writing time for a structure (which smallest planar dimensions are still much smaller than the backscattering length $\beta$) covering a large area (up to a few square millimetres or even larger) should be many orders of magnitude higher for the exposure around the structure than for the exposure of the structure itself. However, for the intrinsic proximity effect suppression discussed here, only a distance of 2-3 times the backscattering length $\beta$ has to be exposed around the desired structure, which brings the writing time of the two possible approaches much closer together. For the same reason it is sufficient to expose only a distance of 2-3 times $\beta$ around the leads to the bond pads (in contrast to figures 2(c) and 3), if a further reduction of the writing time for the electron-beam exposure is desirable, which might be the case for fabricating many devices and/or resists with orders of magnitude higher clearing dose. Now the photolithographically defined leads to the bond pads have to be placed with higher alignment accuracy as shown in figure 3 to prevent a short-circuit.

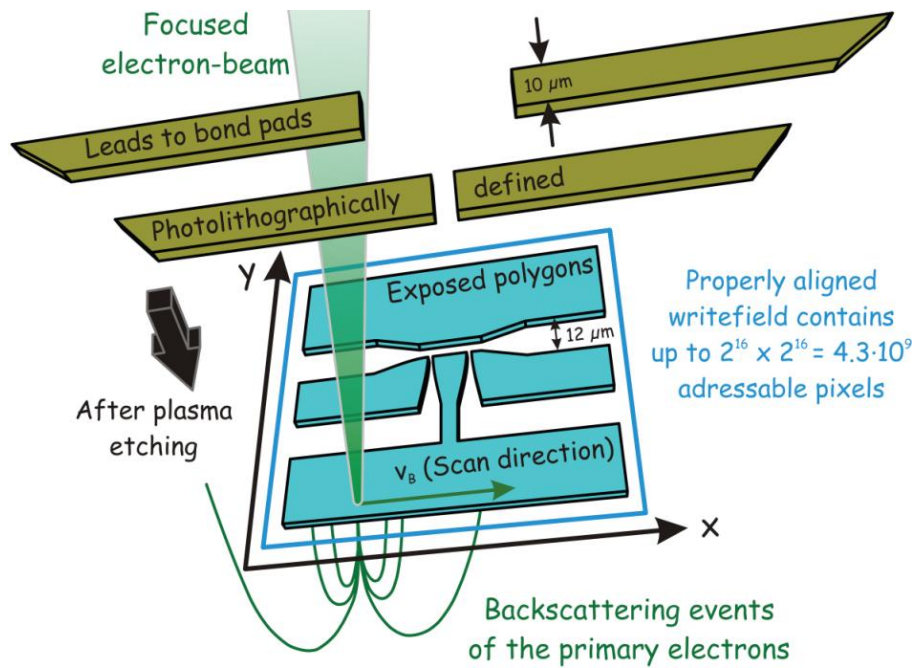

Figure 3: The polygons of a simple bridge located within the write field of the Raith150 are exposed by guiding the focused electron-beam parallel to the $x$-direction. In this case dynamic effects due to acceleration or deceleration of the beam can be neglected as compared to the exposure modes of figure 2. The exposed lines have a spatial pixel separation of $\Delta x=10$nm. The backscattered events are sketched schematically. The designs for the photolithographic mask structures and the nanolithographic polygons are adapted to each other and overlap by one 1µm (according to the shadowing effects observed in figures 7(d) and (e)). After the fabrication of the nanoscaled structure (chapter 2.3), the bond pads and their leads are placed by means of photolithography (see chapter 3).

*2.2.3 Correction for Proximity Exposures for the Design of a Bridge and a SQUID*

Before we compare quantitatively the modelled intrinsic proximity effect suppression approach to nanoscopically measured length scales (see figure 5) fabricated for the case of a nanowire detector design (see figure 2), we now discuss theoretically two fundamentally important designs: a bridge and a SQUID (shown in figure 4). The bridge's design finds applications as a simple current path, but it can also be used for instance to define a transistor's top-gate [36]. The SQUID finds broad applications within superconducting electronics [31,32]. The example SQUID design investigated here does not show the required separation of the latterly fabricated superconductor on the order of the coherence length. This separation of the circular current path has to be generated lithographically afterwards which will not be discussed here.

For both planar designs, we will theoretically investigate the variation of the current path's width $w$ (defined in figures 4(g) and (o)) in four steps ($100\text{nm}$, $500\text{nm}$, $2.5\text{μm}$ and $10\text{μm}$). Accordingly the path width is varied from spatial dimensions orders of magnitude smaller than the backscattering length to a multiple of $\beta$. The results of NanoPECS$^{\text{TM}}$, calculated for a sapphire substrate, for the exposure around (figures 4(a)-(d)) and the exposure of (figures 4(e)-(h)) the bridge as well as the exposure around (figures 4(i)-(l)) and the exposure of (figures 4(m)-(p)) the SQUID are shown in figure 4.

For the bridge as well as the SQUID design, the exposure around the structure needs virtually no correction for device dimensions which are significantly smaller than the backscattering length $\beta$ (see figures 4(a),(b),(i),(j)). The situation is entirely different if the complementary design is exposed (see figures 4(e),(f),(m),(n)) and drastic corrections for proximity exposures become necessary. As soon as the designed path width becomes bigger or equal to $\beta$ (see figures 4(c),(d),(g),(h),(k),(l),(o),(p)), the advantage to intrinsically suppress proximity exposures is lost. In case the SQUID's inner diameter would be reduced into the nanoscale, the design of figure

4(p) would serve as the one that intrinsically suppresses proximity exposures at the inner border of the SQUID's current path.

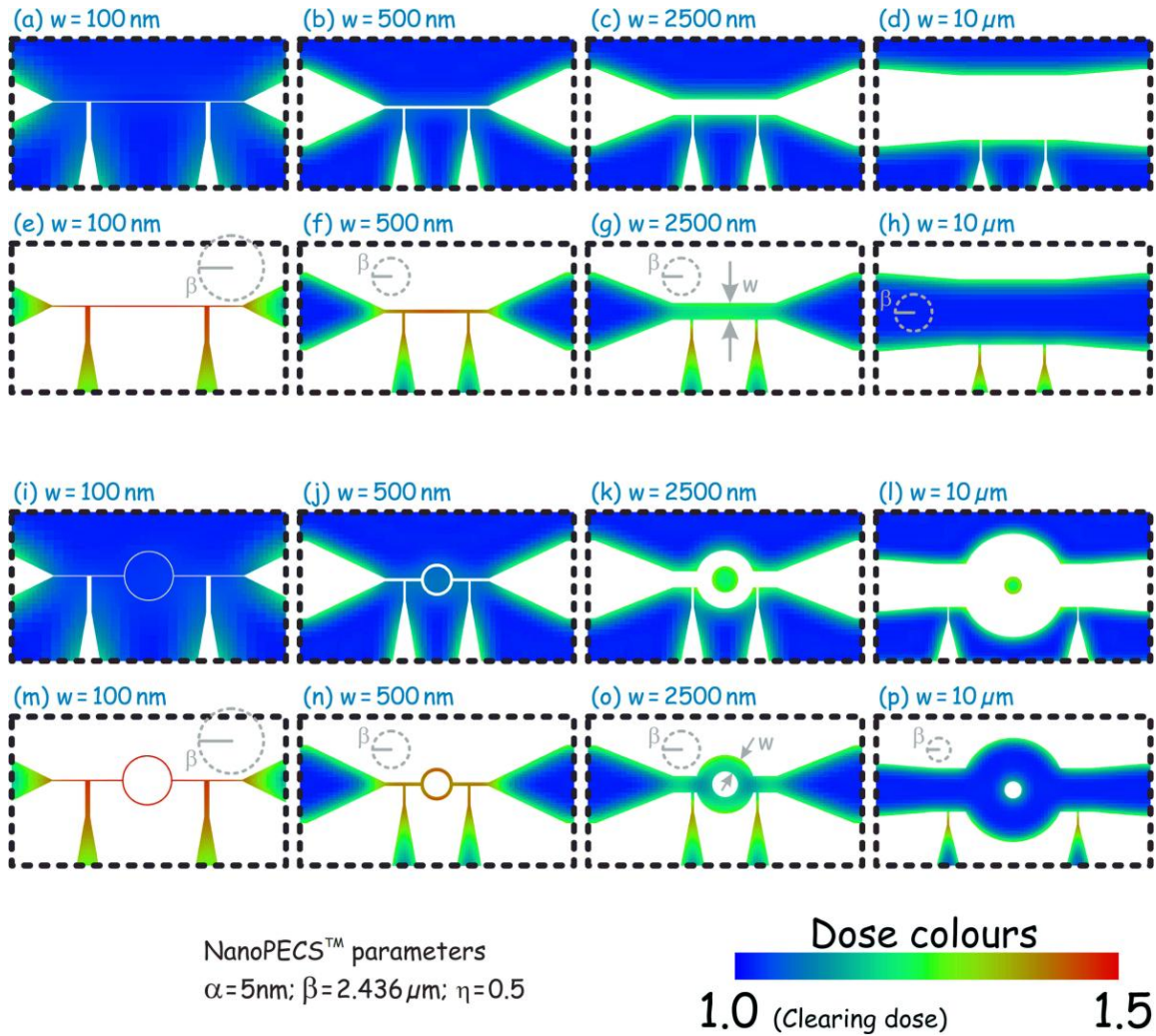

Figure 4: Correction for the proximity effect for exposures around the desired structure ((a)-(d), (i)-(l)) and exposures of the desired structure ((e)-(h), (m)-(p)), calculated by NanoPECS[TM] for the design of a bridge (a)-(h) and for a SQUID (i)-(p) on a sapphire substrate. The 16 planar simulations are scaled to the backscattering length $\beta = 2.436$μm (indicated by the dashed grey circle; this scale is identical for exposures of the desired design and its inverse). The virtual bridge measures 10μm in length. The voltage leads have the same layout as the meander in figure 2. The SQUID's inner radius is 2μm. The distance between the voltage leads in case of the SQUID

is 10μm, for $w<10$μm. For $w=10$μm, the distance between the voltage leads is 30μm. The bridge's as well as the SQUID's current path width is varied in four steps as labelled above the individual simulations. The colours indicate the dose in multiples of the clearing dose value 1.0.

*2.3 Fabrication Results in the Sub-100nm Regime Without Correction for the Proximity Effect*

Our superconducting single-photon detectors [14,23] have been fabricated in form of a nanoscaled meander structured from a cathode-sputtered niobiumnitride (NbN) thin film of thickness $d=5$nm on a sapphire substrate. As compared to the Bethe-range of the electrons which is in the micrometer range for the used sapphire-substrate and an electron acceleration voltage of 30kV, we disregard the influence of the presence of the thin NbN layer on the results discussed in chapter 2.2.

It is well known that niobium containing structures are particularly reactive to a fluoride-based type of chemistry [37-40], and therefore a $SF_6$/Ar plasma discharge was chosen to selectively remove the NbN on the nanoscale. The radio-frequency ($RF=13.56$MHz) plasma was operated at a low microwave power, in order to minimize damages from the ionic bombardment, and at low pressure to ensure vertical bombardment during the etching [41,42]. For the reasons given in chapter 2.2 we decided to use a positive etch protection mask which makes an elaborate correction for proximity exposures for the investigated meander designs redundant. The nanoscaled meander structures discussed below were exposed according to figure 3 while the scanning direction of the beam was adjusted parallel to the conduction paths to prevent a beam-interruption during their fabrication. We have chosen the organic resist ZEP 520A (see part I in [23]) due to its higher chemical robustness and intrinsically higher contrast as compared to PMMA 950k. The higher chemical stability in a reactive plasma allows for a significant reduction of the etch mask height as compared to PMMA. This reduces the spatial scale where forward scattering takes place and hence allows for narrower critical dimensions (see figure 1 in [43]). We precisely determined the etching rates of the material NbN and the organic etching mask (NbN: 1.2nm/min; ZEP 520A: 17nm/min, both at $V_{DC} = -70V$ sheath plasma voltage) with an atomic force microscope (Research AFM from *Asylum*). The sapphire was not affected by the low-power discharge and served therefore as an etch stop layer. Consequently, a height of the ZEP 520A of 70nm ensures the protection of the covered 5nm NbN as long as the unprotected parts of the NbN film are removed. Sub $-100$nm thin resist layers show in general a relatively less pronounced proximity effect. This is a direct consequence of the short path that a highly energetic elastically backscattered electron travels through the ultra-thin resist layer, which makes it less probable that the backscattered electron interacts with the resist before it leaves the sample [44]. In order to increase its intrinsic contrast even more, we chemically developed the exposed ZEP 520A in *n*-Amylacetate at $-10°C$ [45] for 60s. Under these developing conditions, with 70nm high ZEP 520A and 30kV electron acceleration voltage, the organic resist has a clearing dose value of $55\mu C/cm^2$. The whole chip was in focus of the scanning beam by defining a focal plane with the well-known contamination dot technique. A detailed discussion about an approach to properly operate the RF-plasma discharge to obtain a smooth, selective, sensitive and controlled thin film ablation can be found in part I of [23].

The insulating character of the sapphire makes it difficult to precisely measure the geometric dimensions of the fabricated structures with a scanning electron microscope (SEM). Therefore three clones of every of the above mentioned four detector designs with different conduction path widths were lithographically etched 8nm into a 10nm NbN film. The remaining 2nm NbN on top of the substrate ensures the drain of the electrons from the scanning-beam during electrography (we used an ULTRA55 from *Zeiss*) thereby allowing for the investigation of the meanders with the SEM. In figure 5 we show an example electrograph of the explained approach after etching and stripping of the resist residuals.

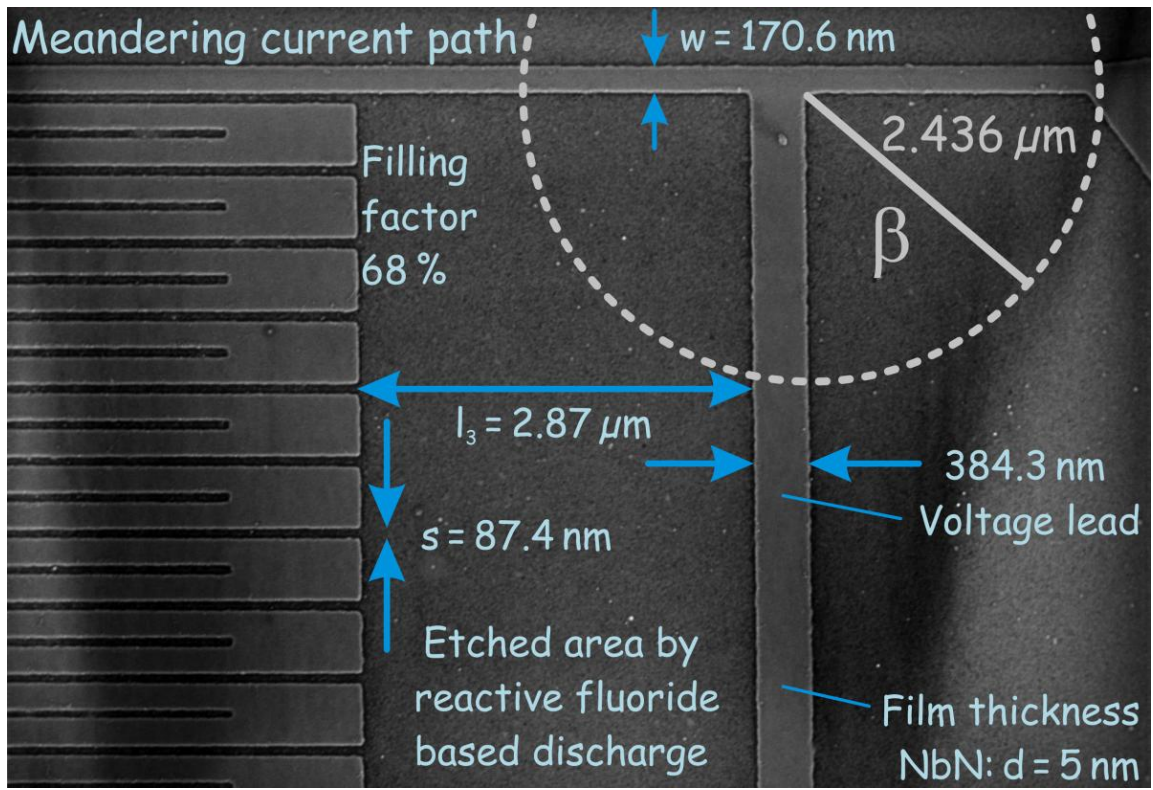

Figure 5: Electrograph of the etched nanoscaled meander that was protected by the positive organic resist ZEP 520A analogous to the EBL layout shown in figure 2. The chemical development and the dry etching slightly modify the design values $w=200\text{nm}$, $s=60\text{nm}$. The structure was not corrected for proximity effects during the guided scanning of the electron-beam during the exposure, according to figure 2 (exposure mode analogous to figure 3).

The chemical development of the ZEP 520A and the dry etching lead to a reduction of the conduction path width by about $30\text{nm}$. This discrepancy for the width $w$ between the design value (see figure 2) and the SEM measured values (see figure 5) was indeed observed on every investigated structure and its clones. The geometry of the meanders with designed conduction path widths $90\text{nm}$, $120\text{nm}$, $200\text{nm}$ and $300\text{nm}$ was measured several times at three different locations spread over the meanders' area. The averaged values from nine SEM measurements were $w=53.4\text{nm}\pm2.3\text{nm}$; $w=82.9\text{nm}\pm2.2\text{nm}$; $w=170.6\text{nm}\pm2.6\text{nm}$; and $w=273.1\text{nm}\pm2.1\text{nm}$. The relative reduction for the larger relevant lengths $(l_1, l_2, l_3, a_I \text{ and } b_I)$ was even smaller (they also shrink by $\cong 30\text{nm}$, which is negligible in this case).

Our fabrication results (without NanoPECS[TM] for the designed path width $w=120\text{nm}$) are directly comparable to those of reference [24] where an elaborate correction for the proximity-effect was performed. The use of a negative organic etch protection mask HSQ (thickness $\cong 70\text{nm}$), required a correction for proximity exposures [24], according to figures 2(e) and (f) of this paper. The path widths obtained by [24] were $w=91.6\text{nm}\pm1.8\text{nm}$ with NanoPECS[TM] and $w=90.1\text{nm}\pm4.3\text{nm}$ without correcting for the proximity effect (we obtained these values by digitizing the figure 2(d) from [24]). Recapitulatory we conclude that our *intrinsic* proximity suppression approach without NanoPECS[TM] reaches the same patterning accuracy of about $\pm 2\text{nm}$ as [24] with NanoPECS[TM]. Even without a complex and time-consuming correction for the proximity effect, we reach the same fabrication accuracy as [24] due to the advantages of the exposure around the structure (compare figures 2(b),(c) and 3).

Finally, our SEM investigation revealed that our fabrication procedure has a minimum possible design spacing $s=60\text{nm}$ for the conduction paths which increases to $\sim 90\text{nm}$ after passing the nanolithographic fabrication process (see figure 5). A smaller design value for $s$, leads to the merging of the conduction paths after the dry etching (not shown). The SEM measurements translate into the following values of the filling factor $FF=40\%, 51\%, 68\% \text{ and } 78\%$. The difference between the designed and the measured values becomes more pronounced for narrower path widths. Note that for detector applications, a high filling factor is desirable because it increases the absorption probability and hence the quantum efficiency of the single-pixel meander detector (see chapter 13 in part II of [23]).

## 3. Considerations for Contact Photolithography with Sub-µm Resolution

In order to minimize the writing time of the EBL, relatively large structures that are necessary for the calibration of the piezoelectronically controlled laser-interferometer stage are fabricated in most R&D cases with a lift-off deposition technique. The deposition is generated by a standard sputter or evaporation process. Examples for such structures are global and local markers, cutting markers or bond pad leads that connect the nanoscale structure to the macroscopic electrical connections of the measurement setup. These structures have spatial dimensions on the microscale, and in many R&D projects they are defined by optical contact lithography.

For contact photolithography, the minimum fabricational feature-size $\delta x$ is given by (see [6] and equation (1.6) in [1])

$$\delta x = \frac{3}{2}\sqrt{\lambda(g + \frac{1}{2}H_R)}. \tag{3}$$

Here, $\lambda$ is the wavelength of the exposure light, $g$ the gap between mask and photoresist and $H_R$ the thickness of the spin-coated resist. Accordingly, if one wishes to fabricate critical dimensions on the order of the wavelength of the used light source, mask and resist should be in perfect contact. To come as close to this condition as possible, a lithographic removal of the edge bead accumulated during the spin coating procedure at the chip edge is necessary (see pages 17 and 18 in [1]). Because it does not allow for further photolithographic processing after the removal of the edge bead, negative resist is not used throughout this work. To determine quantitatively how much of the resist's border has to be removed to obtain a homogenously flat spin-coated surface after the bead-removal, the well-known Newton's interference rings technique is practicable and well suited as it is a measure for the resist (in)homogeneity. In addition a mask aligner with an automated wedge-error correction as the used MA6 from *Karl Süss* (equipped with a HBO 350 W lamp from *Osram*) is desirable.

The following parts of this chapter focuses on the engineering considerations based on the conclusions resulting from equation (3) for fabricating pattern in the $\text{sub}-\mu\text{m}$ regime with a contact photolithographic mask aligner that generates its UV-photons (Hg i-line, $\lambda \approx 365\text{nm}$) by spontaneous photoemission within a mercury direct-current (DC) plasma arc discharge [13].

*3.1 Choice of the "Right" Resist*

In most R&D tasks, the above mentioned necessary structures (markers, etc.) for an easy and effective mix and match process cover only a small fraction of the chip surface area and are supplied by a lift-off deposition process. A comfortable fabrication of these structures during the lift-off procedure is especially enabled by the hardening of a top-layer in chlorobenzene of the positive ®AZ-type photosensitive resist [46]. This resist develops a sharp pronounced T-profile that eases the lift-off procedure even if the deposition was angle-generated. It allows in addition for the removal of the edge-bead to achieve the resolution predicted by equation (3). Unfortunately, the positive character of these ®AZ-type resists translates into a predominantly non-transparent photolithographic mask with mainly black chromium structures on top. This makes the photolithographic alignment procedure complicated because it is difficult to distinguish existing alignment structures on the chip surface during the mask alignment when most areas of the photolithographic mask are black. Fortunately, when working with a negative or image reversal resist, the mask layout has to be inverted for generating an identical pattern. In this case the whole mask is basically transparent. In combination with the above discussed advantages to remove the edge bead, we have therefore chosen the image reversal resist AZ5214E (spin-coated to a height of $H_R = 1.3\mu\text{m}$, measured with the standard surface step profiler Alpha-Step 500 from *Tencor*) for the photolithographic supply of the mix and match structures whose process chain is sketched in figure 6.

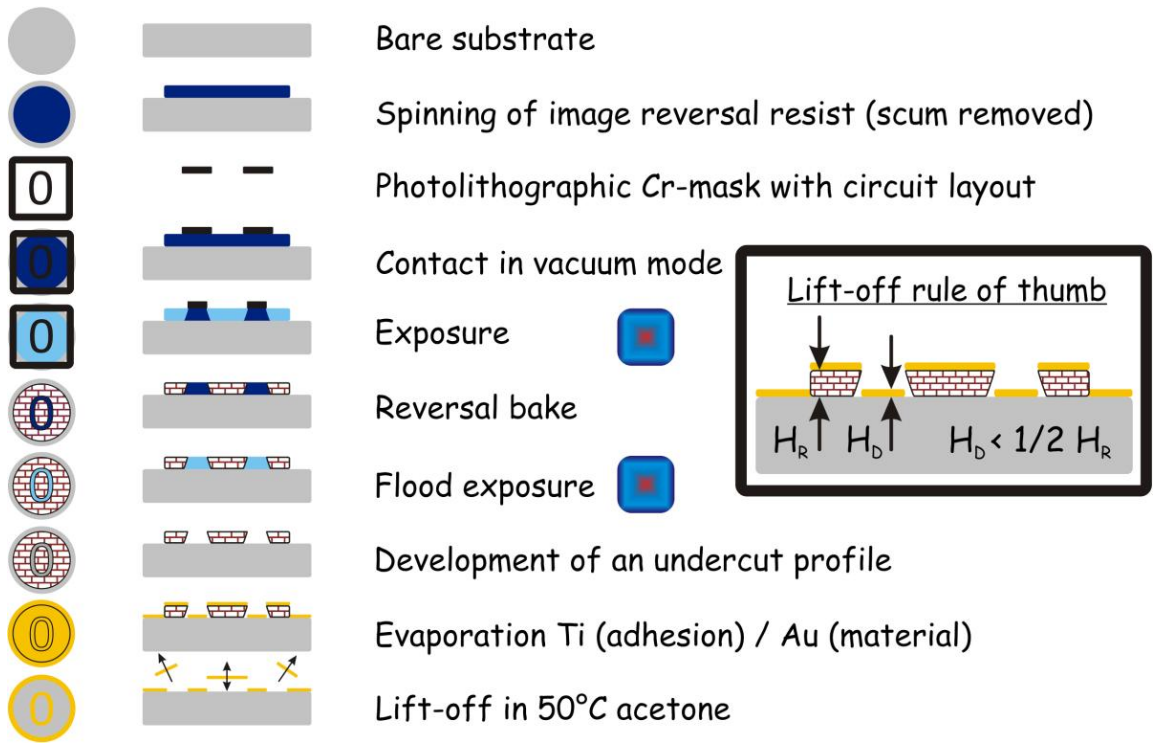

Figure 6: Schematic process chain that is required for a successful *in situ* patterning of an evaporated metallization with an image reversal resist. Details are described in the text. SEM pictures of the last three steps are shown in figure 7. The character "O" was chosen as an example for an enclosed structure. The inner part of the zero is critical during the lift-off process because it is separated from the rest of the film and might contact the chip surface. In the worst case this would cause a short circuit. Therefore a special scriptum for device labelling was used (see figure 9). This scriptum has no enclosed structures and is therefore suitable for a successful lift-off process. Note that none of our photolithographic masks contains any enclosed structures to ease a lift-off process.

During contact photolithography the resist receives the exposure dose $21\,\text{mJ/cm}^2$ with respect to the Hg i-line $(\lambda \cong 365\,\text{nm})$. The photons spontaneously emitted during the DC-discharge in the UV-lamp are absorbed with exponentially decreasing probability by the photoactive components inside the photosensitive resist due to the Lambert-Beer law [47]. Afterwards, the sample is re-versal baked for two minutes at a temperature of 115°C. During this procedure the previously exposed organic polymers hardens and becomes insoluble in the developing chemical *and* insensitive to further UV-light illumination, which is visualized by the brick pattern in figure 6. To chemically develop a binary resist pattern on top of the sample, the whole chip is then exposed to a one order of magnitude higher ultra-violet light dose as compared to the exposure step (flood exposure $140 mJ/cm^2$, again adapted to the i-line). Now the previously unexposed parts become soluble while the previously exposed parts are still insoluble in the chemical developer. As a consequence of the exponential absorbance of the exposure dose inside the resist, the top resist layer is now less soluble in the chemical developer (MF 319 from *Shipley*) as the resist layer closer to the substrate surface. The development for 1min and 35s reveals, therefore, the necessary undercut profile (see figure 7(b)) to lift-off an evaporated or sputtered metallization layer. After the deposition, the chip is placed in a bath of acetone kept at 50°C. The acetone dissolves the organic resist and percolates underneath the evaporated metallization which is not in contact with the substrate. We have chosen the process parameters according to the product data sheet from *Clariant* [48] and optimized them for the here used equipment until the results of figure 7 were obtained.

*3.2 Critical Dimensions & Resist Profile*

In this section we discuss the critical dimensions that we achieved with the straightforward application of the above explained principles. The results discussed below were obtained with a 3" silicon wafer. The borders of such a commercial wafer are rounded in order to reduce the component of the surface tension pointing towards the axis of rotation, which reduces the height of the edge bead. After the spin-coating procedure, we lithographically removed 5mm of resist close the wafer borderline to ensure a nearly perfect contact between mask and resist during the expo-sure. After the chemical development and illumination with a yellow-filtered light source no Newtonian interference ring was observed under the optical microscope. The last three fabrication steps of the patterning chain shown schematically in figure 6 were investigated by applying a standard cleaving procedure. Our motivation for this examination was to observe the actual shape of the undercut profile which forms during the chemical development according to the decreasing absorbance within the resist layer during the UV-light exposure step. A series of fine straight lines (see chapter 3.3, figure 8) were fabricated with varying exposure dose adapted to the Hg i-line as labelled in figures 7(a)-(c). The black chromium structures on the photolithographic mask for the investigation in figure 7 were 1μm wide (respectively 0.5μm, see figure 7(f)), 100nm thick and spaced 1μm (respectively 0.5μm, see figure 7(f)) apart from each other. After the chemical development and the cleaving of the samples, a thin conductive film (2nm gold) was sputtered on top to suppress charging effects during determining the undercut profile with the SEM (see figures 7(a)-(c)).

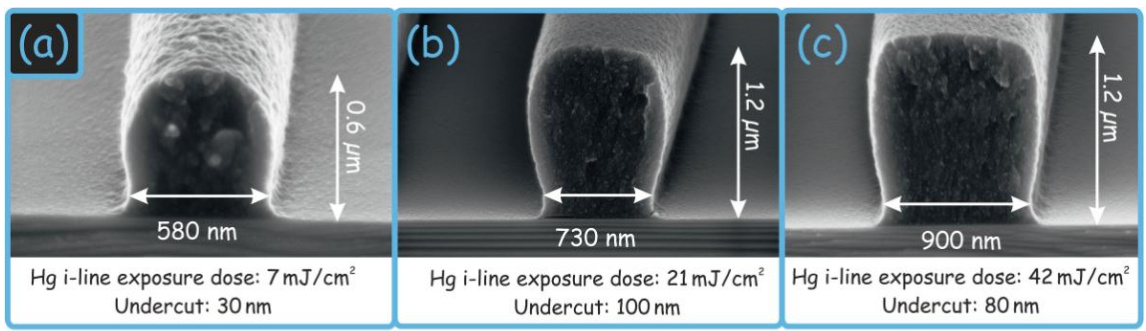

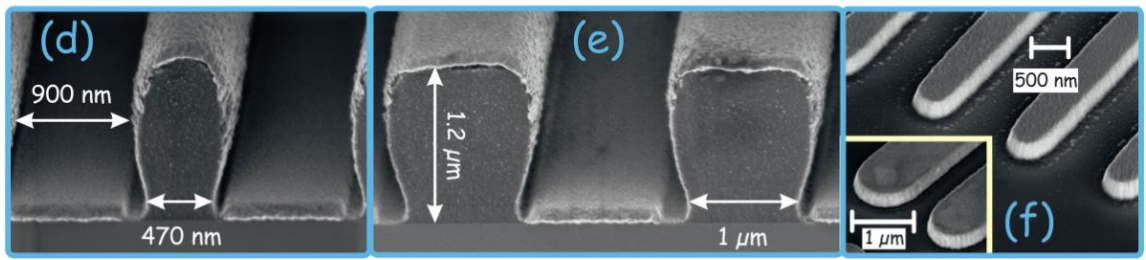

Figure 7: The electrographs (a)-(c) show the results of an exposure dose variation. The Pictures (d) and (e) show the structured organic resist after the evaporation of a 100nm thick metallization layer. Picture (f) illustrates line widths after the lift-off procedure in acetone that are very close to the wavelength of the mercury arc discharge lamp ($\lambda = 365\text{nm}$) as predicted by equation (3). The electrographs were taken with tilted stage, explaining the discrepancy of $0.1\mu\text{m}$ between the height of the organic resist measured within the pictures (b),(c),(e) and the value given in the main text. All structures are located on a silicon substrate to allow for a cleaving inspection with the SEM.

For the fabrication process of devices, the exposure dose was fixed to $21\text{mJ/cm}^2$ (leading to the result of figure 7(b)), because it produced the most pronounced undercut within the investigated experimental parameter frame. Using this optimal dose, and after the evaporation of a titanium/gold metallization layer $(\approx 100\text{nm})$, an identical cleaving test on a different sample was performed (see figures 7(d),(e)). The Ti $(10\text{nm})$ serves as an adhesion layer, and the Au was chosen as the material for the bond pads and marker structures because it is chemically inert, so that rebonding of the same structure is unproblematic, and it is easily recognized during an automated mark recognition procedure due to its higher atomic number as compared to the sapphire and silicon substrate. We used a standard electron-beam evaporation [49,50] tool (Univex 500 from *Leybold*) for the evaporation which measures the condensed height with a quartz oscillating crystal that changes its resonance frequency due to the deposited material during the evaporation that can be measured with an appropriate resonance circuit. The electron-beam evaporator is pumped to a high vacuum on the order of $10^{-6}\text{mbar}$. In this high-vacuum regime, the mean-free path is very large. Hence the evaporated atoms undergo essentially collision less line-of-sight transport prior to condensation on the substrate, thus leading to a thickness build-up directly above the source that decreases slowly away from it. We measured the height of the evaporated structures on a 3" wafer and did not observe any height variation on the wafer scale within the resolution limit of the used surface step profiler ( 1nm [51]).

In figures 7(d),(e) we show the result obtained with 1μm wide chromium mask-structures. They were spaced 500nm, respectively 1μm . Note the very good reproducibility of our photolithographic approach which can be recognized in the identical shapes of the resist profiles in figures 7(b),(e). From the discontinuity of the metallic film one can conclude that the metal vapour scatters very little on its way from the crucible to the chip due to the high-vacuum conditions during the *in situ* condensation. The disconnection of the bottom and top metallic layers is essential for a successful lift-off of the top-layer in the acetone bath. Two further consequences of the high-vacuum conditions are the well defined edges of the evaporated lines and a slight shift of the evaporated metallic lines from the center of the photolithographically opened resist gap (see figures 7(d),(e)). The latter originates from the fact that the sample was not placed above the center of the melted material during the electron-beam evaporation procedure, leading to a slight shadowing effect. Therefore, a sufficient overlap of 1μm between the photolithographically defined pattern and the nanoscaled fabricated devices was included to compensate for that shadowing effect. Therefore the photolithographically defined leads to the bond pads were designed 10μm in width and the EBL layouts 12μm wide (see figure 3).

After the lift-off, the *in situ* structured metallic pattern is on top of the substrate as shown in figure 7(f). These are the smallest structures (500nm in width) that were placed on the photolithographic mask. Taking into account the resist height of 1.3μm , we indeed reach the theoretical resolution limit predicted by equation (3).

*3.3 Controlling the Undercut during Development*

In order to obtain a reasonable control over the undercut during device fabrication, a unique technique that avoids cleaving the wafer was developed which is also applicable for substrates where cleaving is not possible, e.g. for sapphire which was used throughout chapter 2. A matrix of fine straight lines with different widths and spacings is included on the optical chromium mask (see cleaving structures in figure 9). These structures were also used in the last section, where the cross-sectional shape was of interest (see figure 7). Figure 8(a) shows a rendered picture of a section of this cleaving structure matrix included on the chromium mask and located slightly above the transferred pattern by the lift-off deposition process chain of figure 6. Next to each set of straight fine lines two numbers characterize their width and spacing in $\mu m$. In the horizontal direction the line spacing is varied, while in the vertical direction the line width is varied ($0.5\mu m$, $1\mu m$, $2\mu m$, $3\mu m$ and $4\mu m$).

Only one matrix element out of the above defined matrix could not be fabricated with our lift-off deposition process. These were the narrowest $0.5\mu m$ wide lines spaced $0.5\mu m$ (approximately one wavelength of the used mercury light source). The other 24 matrix elements of cleaving structures could be successfully fabricated with the process chain of figure 6 (the situation is rendered in figure 8(a)).

For relatively wide and well separated photo-mask structures, the undercut-profile serves only to ensure a successful lift-off deposition process. However, if the separation of the chromium structures is reduced down to the order of the undercut, the above introduced matrix can be used to estimate the value of the undercut. In figure 8(b) we show one set of straight chromium lines ($3\mu m$ wide, spaced $1\mu m$) out of the above described matrix, the resist pattern and the gold pattern after the lift-off. As soon as the undercut shaping the resist is about half of the spacing between the lines on the mask or larger ($\approx 500nm$ in the case of figure 8(b)), the resist lines ($1\mu m$ in width in the case of figure 8(b)) lose their surface attachment and bend aside which explains their pronounced curvature. Therefore, the undercut after a certain development time can be esti-mated without cleaving the wafer. The presented technique is an excellent control instrument for an existing stable set of process parameters, and should also be straight forward applicable to a nanoscaled lift-off deposition process.

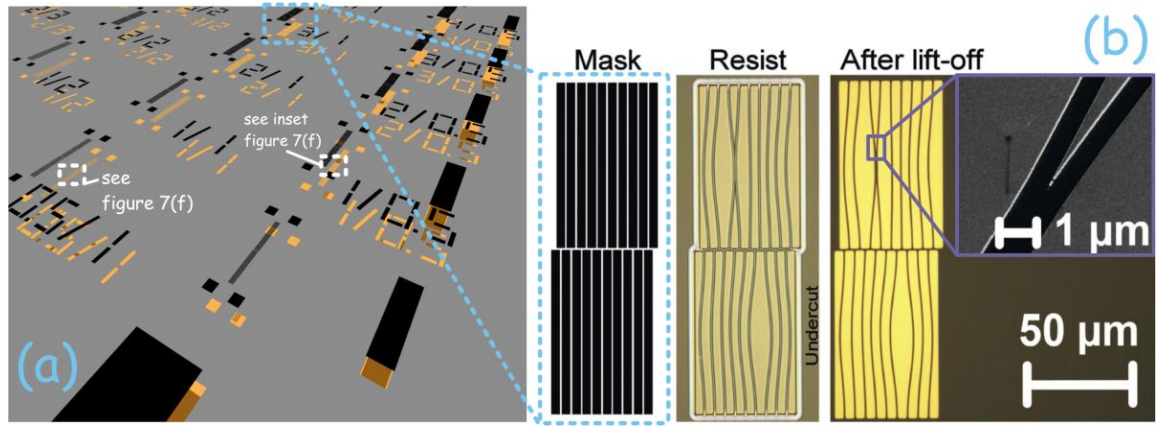

Figure 8: A unique and novel technique was developed in order to determine the undercut of a sample without cleaving it. On the left the rendered picture (a) shows a cut-out of the mask (chromium=black) and the transferred pattern (golden) by the additive patterning process (figure 6). Note that the 0.5μm lines, spaced 0.5μm apart from each other, are located on the chromium mask but were not processable within the process chain of figure 6. The white border displays the location of the critical dimension lines shown in figure 7(f). On the right side (picture (b)) three stages of the pattern transfer are shown from which the undercut can be estimated.

**4. Example for Mix & Match Fabrication Using Photolithography and EBL**

*4.1 Hierarchic GDSII-Design*

A high flexibility that combines the flexible EBL (chapter 2) with the fast photolithography (chapter 3) is achieved by the usage of the hierarchic GDSII (**G**raphic **D**ata **S**ystem) format and different groups of structure references (SRs). The first group is called *coordinate structures* (or-ange in figure 9). All of the following micro- and nanolithographic steps are aligned with respect to this sub−$\mu$m precise coordinate system. Therefore all necessary alignment figures and recognition structures for further processing are placed on this first layer that is fabricated by the above described photolithographic lift-off deposition process (see figure 6) on top of the NbN film. The coordinate structures include **cut markers** that separate the different devices and should be designed according to the specifications of the used wafer saw. The cutting markers allow for fabrication of many devices per single fabrication run. **Alignment structures** allow for the precise ($\approx$100nm accuracy) matching of additional photolithographic steps. **Global** and **local markers** (described above) are necessary for a successful EBL exposure with nanoscaled placement accuracy.

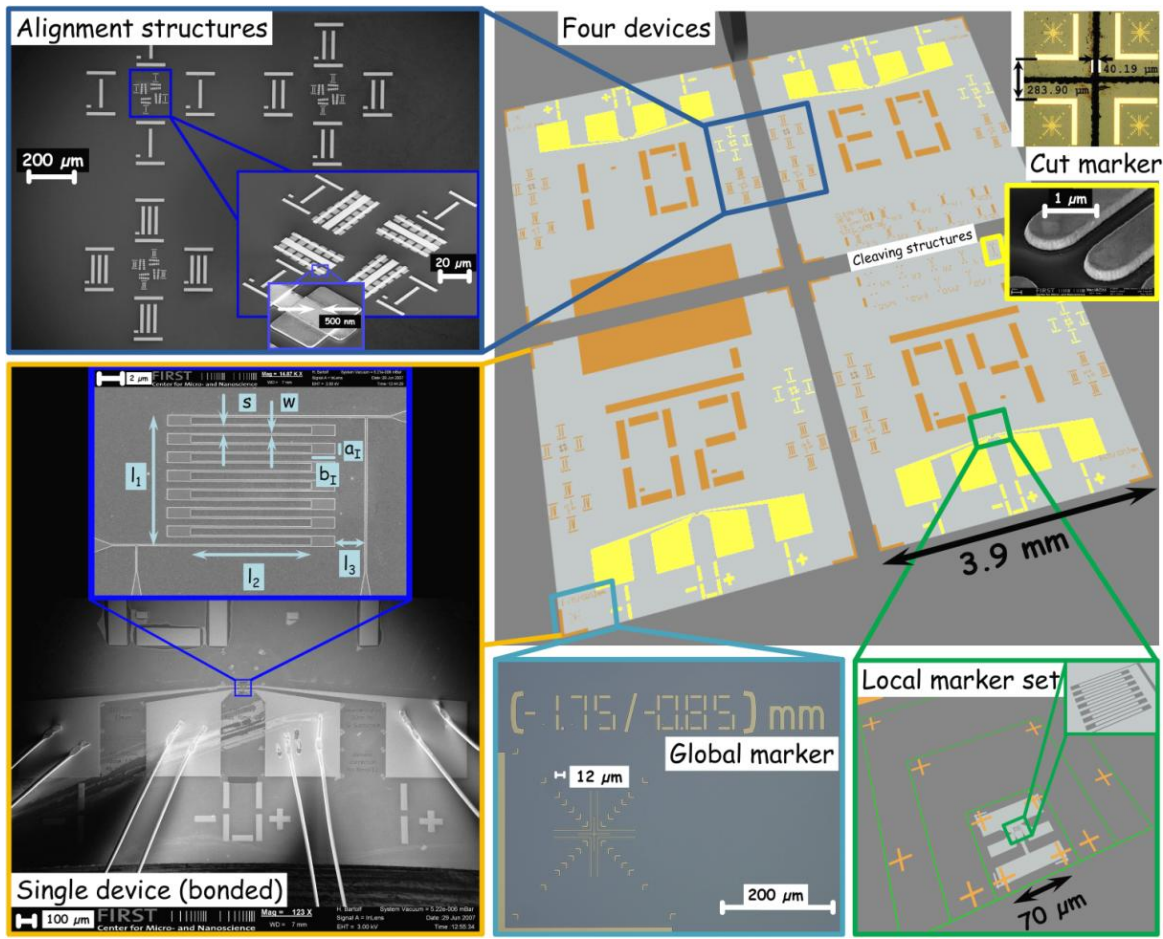

Figure 9: Basic layout of a chip for the fabrication of four devices. The nanoscale definition of the device is created by EBL. The positioning marks and the connection of the nanostructure to the outer world (bond pads via Au-leads) are fabricated by a photolithographic lift-off deposition process. Every electron-beam written structure can be designed individually, and the process layout allows for the definitions of differently sized write fields from 100μm up to 500μm. The displayed four devices are fabricated in one fabrication run and are afterwards separated with a wafer saw.

A second group of SRs (yellow in figure 9) contains the mask data for the photolithographically defined structures (bond pads, wave guides, device protection area etc). The structures for every photolithographic step should be designed on a separate GDSII-layer to facilitate the extraction of the data for the optical mask(s). The rectangles of the alignment structures for each following step must fit the squares of the coordinate step (see inset *"Alignment structures"* of figure 9; and chapter 5 in part I of [23] for a detailed discussion of their layout dimensions). The third group of SRs contains the polygons for the EBL exposure (see figures 2 - 4) and the scans that allow for the automatic alignment of the write field for the deflected but focused electron-beam. The polygons match the coordinates of the photolithographically defined structures and include a sufficient overlap.

All the mentioned SRs are matrixcopied into a higher-ranking SR, named **Process**. The variable distance of the matrix elements allows for the modification of the device density. For example, if we reduce the bond pads to an area of $10^4 \mu m^2$ it is possible to place over 1000 Devices on a 2" wafer. An example design of the above discussed approach that also served for fabricating the structures used in the project [14] is shown in figure 9.

*4.2 Fabrication-Process Chain*

In the previous chapters the developed micro- and nanotechnological approaches and their limitations for several individual fabrication steps has been discussed. In the following we discuss the sequence of the single steps for fabricating a functioning device and explain why we chose this particular process chain (see figure 10).

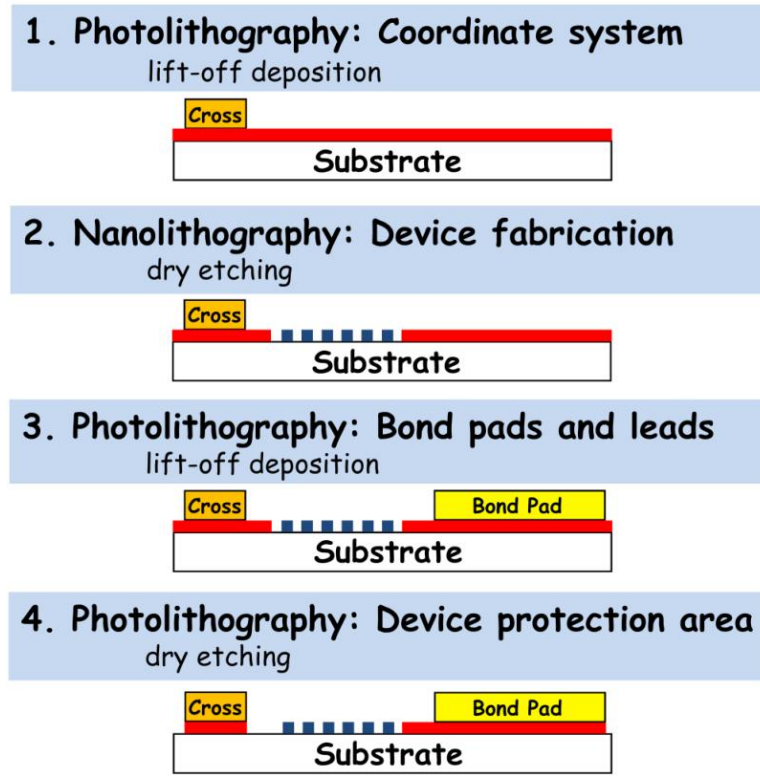

Figure 10: Process chain for the fabrication of pre-sputtered superconducting films. The zeroth step (not shown) is cathode-sputtering of the NbN film (red). The colours are chosen according to figure 9. The NbN nanoscaled pattern is drawn blue.

The *coordinate structures* described in the last chapter were first placed on top of the NbN thin film to allow for the calibration of the piezoelectronically controlled laser-interferometer stage. Then the photolithographically defined bond pads were fabricated after the nanoscale lithography to ensure a homogeneous height of the spin-coated resist within the write-field area for the deflected electron-beam during the EBL exposure. The usage of a positive etch protection mask ZEP 520A during nanolithography has a further advantage, in addition to the already extensively discussed intrinsic proximity exposure suppression of chapter 2.2: After the nanoscale fabrication, only a tiny fraction of the NbN film around the fabricated structure is removed during the dry etching step. According to the design-layout (see left picture series in figure 2), the whole NbN film is interconnected after the dry etching. This is necessary to minimize the contact resistance between the NbN and the evaporated Ti/Au layer of the bond pads placed on top. This contact resistance problem is more difficult to solve with a negative EBL resist. A last photolithographic step is finally necessary that allows for the removal of the short circuited NbN film around the bond pads. During this last dry etching step, the bond pads serve as a protection layer for the NbN underneath. For this step we use the photosensitive resist AZ6632 to protect the write field of the electron-beam. By removing the edge-bead, we achieve critical dimension of 1μm with this resist that we spin-coated to a height of 3μm (see appendix of [23]).

## 5. Conclusion

We have demonstrated and discussed a technique that intrinsically suppresses proximity exposures becoming noticeable at a high electron acceleration voltage during EBL that does not require a proximity correction algorithm. Our approach allows, among other things, for a relatively easy change of the substrate material and it is applicable for device-layout dimensions that are significantly smaller than the backscattering length $\beta$ at the operational high voltage. Furthermore it simplifies the EBL-fabrication procedure especially in case of complicated curved planar device-designs where a proximity correction algorithm is time- and/or computational-power consuming. We could show that our approach is quantitatively very close to the published work of J. K. W. Yang *et al.* [24], who performed an elaborated correction for the proximity effect.

In order to reduce the writing time of the used electron-beam lithograph we developed a fabrication process that allows for extending the optical contact lithography with a mercury arc dis-charge lamp into the $\text{sub}-\mu\text{m}$ range, simply by removing the edge-bead that we checked quantitatively with the Newton's interference rings technique. We also developed a novel technique that allows for a comfortable control of the undercut during a lift-off fabrication procedure.

Finally we discussed the order of the fabrication steps on an example top-down fabrication approach that mixes and matches EBL and photolithography.

**Acknowledgements**

We thank the group of M. Siegel from the Institute for Micro- and Nanoelectronic systems at the University of Karlsruhe for sputtering the thin NbN films for us. This work was partially supported by the Swiss National Science Foundation program NCCR **Ma**terials with **N**ovel **E**lectronic **P**roperties (MaNEP). H. B. acknowledges the Research Funding 2010 from the University of Zürich. The micro- and nanostructuring was carried out at the FIRST Center for Micro- & Nanoscience of ETH Zürich. One of us (H. B.) is thankful to his colleagues Hongang Liu from the Institute of Microelectronics of the Chinese Academy of Sciences in Beijing, Andreas Alt from the Laboratory for Millimeter-Wave Electronics of the ETH Zürich and Patric Strasser from the Electronics Laboratory of the ETH Zürich for stimulating discussions concerning the physical impact of the clean-room equipment parameters used in the FIRST laboratory. We are thankful to Guido Piaszenski and the team from Raith GmbH in Dortmund for a successful and fruitful collaboration. Throughout this paper the international SI unit system was used.

References

[1]   Mack C 2007 *Fundamental Principles of Optical Lithography: The Science of Microfabrication* (John Wiley & Sons, New Jersey, United States of America)

[2]   Brewer R G 1980 *Electron Beam Technology in Microelectronic Fabrication*, ed R G Brewer (Academic Press, San Diego, United States of America)

[3]     Tseng A A, Chen K, Chen C D, Ma K J 2003 *IEEE Trans. Elect. Pack. Manufac.* **26** 141

[4]     Reimer L 1998 *Scanning Electron Microscopy* (Springer, Berlin, Germany)

[5]     Goodhew P J, Humphreys J, Beanland R 2001 *Electron Microscopy and Analysis* (Taylor & Francis, London, United Kingdom)

[6]     Mailly D 2009 *Eur. Phys J. Special Topics* **172** 333

[7]     Seiler H 1983 *Appl. Phys.* **54** R1

[8]     Chang T H P 1975 *J. Vac. Sci. Technol.* **12** 1271

[9]     Wüest R, Strasser P, Jungo M, Robin F, Erni D, Jäckel H 2003 *Microelec. Engineer.* **67-68** 182

[10]    Robin F, Costea S, Stark G, Wüest R, Strasser P, Jäckel H, Rampe A, Levermann M, Piaszenski G 2005 *Accurate Proximity-Effect Correction of Nanoscale Structures with NanoPECS* (Raith Application Note, Technical Report)

[11]    Greeneich J S 1980 *Electron Beam Technology in Microelectronic Fabrication* [2], ed R G Brewer (Academic Press, San Diego, United States of America) chapter 2 p 66

[12]    http://www.first.ethz.ch/infrastructure/equipment/EBL_Raith150Spec.pdf (June 2010)

[13]    Lister G G, Lawler J E, Lapatovich W P, Godyak V A 2004 *Rev. Mod. Phys.* **76-68** 541

[14]    Bartolf H, Engel A, Schilling A, Il'in K, Siegel M, Hübers H-W, and Semenov A 2010 *Phys. Rev. B* **81** 024502; Selected for the January 15, 2010 issue of the Virtual Journal of Applications of Superconductivity

[15]    Engel A, Bartolf H, Schilling A, Semenov A, Hübers H-W, Il'in K and Siegel M 2009 *J. Mod. Opt.* **56** 352

[16]    Semenov A D, Gol'tsman G N, Korneev A A 2001 *Physica C* **444** 12

[17]    Gol'tsman G N, Okunev O, Chulkova G, Lipatov A, Semenov A, Smirnov K, Voronov B, Dzardanov A, Williams C, Sobolewski R, 2001 *Appl. Phys. Lett.* **79** 705

[18]    Jaspan M A, Habif J L, Hadfield R H, Nam S W 2006 *Appl. Phys. Lett.* **89** 031112

[19]   Hadfield R H, Habif J L, Schlafer J, Schwall R E, Nam S W 2006 *Appl. Phys. Lett.* **89** 241129

[20]   Honjo T, Nam S W, Takesue H, Zhang Q, Kamada H, Nishida Y, Tadanaga O, Asobe M, Baek B, Hadfield R, Miki S, Fujiwara M, Sasaki M, Wang Z, Inoue K, Yamamoto Y 2008 *Opt. Express* **16** 19118

[21]   Kyser D F, Viswanathan N S 1975 *J. Vac. Sci. Technol.* **12** 1305

[22]   Ivin V V, Silakov M V, Vorotnikova N V, Resnick D J, Nordquist K N, Siragusa L 2001 *Microelec. Engineer.* **57-58** 355

[23]   Bartolf H 2010 *Fabrication and Characterization of Superconducting Nanowire Highspeed Single-Photon Detectors* (Ph.D. thesis, MNF faculty, University of Zürich, Switzerland)

[24]   Yang J K W, Dauler E, Ferri A, Pearlman A, Verevkin A, Gol'tsman G, Voronov B, Sobolewski R, Keicher W E, Berggren K K 2005 *IEEE Trans. Appl. Supercond.* **15** 626

[25]   Kratschmer E 1981 *J. Vac. Sci. Technol.* **19** 1264

[26]   Stevens L, Jonckheere R, Froyen E, Decoutere S, Lanneer D 1986 *Microelec. Engineer.* **5** 141

[27]   Patrick W, Vettiger P 1988 *J. Vac. Sci. Technol. B* **6** 2037

[28]   Dubonos S V, Gaifullin B N, Raith H F, Svintsov A A, Zaitsev S I 1993 *Microelec. Engineer.* **21** 293

[29]   Dix C, Flavin P G, Hendy P, Jones M E 1985 *J. Vac. Sci. Technol. B* **3** 131

[30]   Parikh M 1979 *J. Appl. Phys.* **50** 4371; 1979 *J. Appl. Phys.* **50** 4378; 1979 *J. Appl. Phys.* **50** 4383; 1980 *J. Appl. Phys.* **51** 700; 1980 *J. Appl. Phys.* **51** 705

[31]   Clarke J, Braginski A I 2004 *The SQUID Handbook: Volume I: Fundamentals and Technology of SQUIDs and SQUID Systems* (Wiley-VCH Verlag GmbH & Co. KGaA, Weinheim, Germany);

[32]     Clarke J, Braginski A I 2006 *The SQUID Handbook: Volume II: Applications of SQUIDs and SQUID Systems* (Wiley-VCH Verlag GmbH & Co. KGaA, Weinheim, Germany)

[33]     Hovington P, Drouin D, Gauvin R 1997 *Scanning* **19** 1

[34]     Drouin D, Hovington P, Gauvin R 1997 *Scanning* **19** 20

[35]     Hovington P, Drouin D, Gauvin R, Joy D C, Evans N 1997 *Scanning* **19** 29

[36]     Locquet J P, Marchiori C, Sousa M, Fompeyrine J, Seo J W 2006 *J. Appl. Phys.* **100** 051610

[37]     Picard A, Turban G 1985 *Plas. Chem. Plas. Proc.* **5** 333

[38]     Lichtenberger A W, Lea D M, Lloyd F L 1993 *IEEE Trans. Appl. Supercond.* **3** 2191

[39]     Williams K R, Muller R S 1996 *J. Micromech. Sys.* **5** 256

[40]     Williams K R, Gupta K, Wasilik M 2003 *J. Micromech. Sys.* **12** 761

[41]     Vossen J L, Cuomo J J 1978 *Thin Film Processes I*, ed J L Vossen and W Kern (Academic Press, San Diego, United States of America) chapter *Glow Discharge Sputter Deposition* p 11

[42]     Rossnagel S M 1991 *Thin Film Processes II*, ed J L Vossen and W Kern (Academic Press, San Diego, United States of America) chapter *Glow Discharge Plasmas and Sources for Etching and Deposition* p 11

[43]     Bartolf H, Engel A, Schilling A, Il'in K, Siegel M 2008 *Physica C* **468** 793

[44]     G. Piaszenski, Raith GmbH Dortmund, private communication

[45]     Ocola L E, Stein A 2006 *J. Vac. Sci. Technol. B* **24** 3061

[46]     Hatzakis M, Canavello B J, Shaw J M 1980 *IBM J. Res. Develop.* **24** 452

[47]     Beer A 1852 *Ann. d. Phys.* **86** 78

[48]     http://groups.mrl.uiuc.edu/dvh/pdf/AZ5214E.pdf  (June 2010)

[49]     SreeHarsha K S 2006 *Principles of Physical Vapor Deposition of Thin Films* (Elsevier Science & Technology, San Diego, United States of America)

[50]     Deshpandey C V, Bunshah R F 1978 *Thin Film Processes II*, ed J L Vossen and W Kern (Academic Press, San Diego, United States of America) chapter *Evaporation Processes* p 79

[51]     http://www.first.ethz.ch/infrastructure/equipment/ssp (June 2010)